 \documentclass[12pt,preprint]{aastex}

\slugcomment{Not to appear in Nonlearned J., 45.}

\shorttitle{Collapsed Cores in Globular Clusters}
\shortauthors{Djorgovski et al.}

\begin{document}


\title{No evidence for a dependence of the mass--size relation of early-type galaxies on environment in the local Universe}


\author{M. Huertas-Company\altaffilmark{1}, F. Shankar\altaffilmark{1}, S. Mei\altaffilmark{1}, M. Bernardi\altaffilmark{2}, J.A.L. Aguerri\altaffilmark{3}, A. Meert\altaffilmark{2}, V. Vikram\altaffilmark{2}}
\affil{GEPI - Paris Observatory, France}
\affil{Department of Physics and Astronomy, University of Pennsylvania, Philadelphia, PA 19104, USA}
\affil{Instituto de Astrofisica de Canarias, E-38200 La Laguna, Tenerife, Spain}


\altaffiltext{1}{University of Paris Diderot}
\altaffiltext{3}{Departamento de Astrofisica, Universidad La Laguna, E-38205 La Laguna, Tenerife, Spain}


\begin{abstract}
   The early--type galaxy (ETG) mass--size relation has been largely studied to understand how these galaxies have assembled their mass. One key observational result of the last years is that massive galaxies increased their size by a factor of a few at fixed stellar mass from $z\sim2$. Hierarchical models favor minor mergers as a plausible driver of this size growth. Some of these models, predict a significant environmental dependence in the sense that galaxies residing in more massive halos tend to be larger than galaxies in lower mass halos, at fixed stellar mass and redshift.  %This environmental dependence is predicted to be stronger for central galaxies.  \\
  At present, observational results of this environmental dependence have been contradictory. In this paper we revisit this issue in the local Universe, by investigating how the sizes of massive ETGs depend on large-scale environment using an updated and accurate sample of ETGs in different environments - field, group, clusters - from the Sloan Digital Sky Survey DR7.   \\
%We select ETGs based on the automated morphological classification by Huertas-Company et al. (2011), and use galaxy sizes from 2D sersic model fitting performed by Meert et al. (2012). Our group and cluster sample is from Yang et al. (2007), updated to the DR7. We compare our results to predictions from the hierarchical models from Shankar et al. (2011) and Guo et al. (2010). \\
Our analysis does not show any significant environmental dependence of the sizes of central and satellites ETGs at fixed stellar mass at $z\sim0$. The size-mass relation of early-type galaxies at $z\sim0$ seems to be universal,  i.e., independent of the mass of the host halo and of the position of the galaxy in that halo (central or satellite). The result is robust to different galaxy selections based on star formation, morphology or central density. In fact, considering our observational errors and the size of the sample, any size ratio larger than $30-40\%$ between massive galaxies ($log(M_*/M_\odot)>11$) living in clusters and in the field can be ruled out at $3\sigma$ level. 

%We compare our observational results with two hierarchical models built from the Millennium Simulation. Models tend to predict a stronger signal with environment than observed. However, once observational errors are properly convolved with model predictions, the tension is erased at least for one of the models.  {\bf The theoretical implications of this result are carefully discussed in a companion paper.}

\end{abstract}



\section{Introduction}
\label{sec:intro}
The study of scaling relations at low and high redshift (e.g \citealp{2010MNRAS.404.2087B, 2011MNRAS.412L...6B,2010MNRAS.405..948S}) is a powerful tool to constrain models of galaxy evolution. In particular, the mass--size relation has been largely studied in the recent literature. One key observational result arising from many of these works is that massive galaxies experienced a strong size evolution in the last 10 Gyrs , e.g., a significant fraction of them increased their size by a factor 2-3 from $z\sim1$ and by $3\sim5$ from $z\sim2$ (e.g. \citealp{2005ApJ...626..680D, 2006ApJ...650...18T,2008ApJ...688...48V,2008ApJ...677L...5V,2008ApJ...687L..61B,2011ApJ...739L..44D,2012MNRAS.422L..62C,2012MNRAS.tmp..141H,2012ApJ...745..130R,2012ApJ...754..141M}).
%These first results have been confirmed by recent measurements of the dynamical masses of these compact high--redshift objects (e.g. \citealp{2011ApJ...738L..22M, 2011ApJ...736L...9V, 2010ApJ...717L.103N}) .

Models of galaxy formation have proposed two main mechanisms to increase the size of early-type galaxies (ETGs). \cite{2008ApJ...689L.101F} proposed mass loss via AGN feedback as the main process responsible for galaxy expansion (expansion scenario) while \cite{2009ApJ...691.1424H} and \cite{2009ApJ...699L.178N} argued that minor dry mergers are the most efficient mechanism (see also \citealp{2011arXiv1105.6043S}). Since both mechanisms act in very different time scales (e.g \citealp{2012MNRAS.423.3243R}) and leave different imprints in the galaxy structure (e.g \citealp{2009ApJ...691.1424H}), these observables have been largely used to constrain the models. Observational evidence clearly supporting one of the above theoretical proposals is still debated in the literature. On the one hand, \cite{2011MNRAS.415.3903T} reported that the low scatter in the ages of galaxies is difficult to reconcile with the fast growth predicted by the expansion scenario and \cite{2010ApJ...709.1018V, 2012arXiv1208.0341P} among others showed that galaxies grow inside-out and increase their sersic index which are clear predictions of the merger models. On the other hand, \cite{2011ApJ...726...69A} for instance, claim significant evolution in size but no in Sersic index for Brightest Cluster Galaxies (BCGs), supporting an expansion scenario rather than a merger-driven one. Also, Newman et al. (2012) reported recently that only if a short dynamical time scale is assumed, mergers alone can explain the growth shown by the data from $z\sim1$(see also \citealp{2012A&A...548A...7L}) and \cite{2012MNRAS.tmp..141H} showed that several state-of-the-art semi analytical models struggle to fully reproduce the size evolution for galaxies at fixed $log(M_*/M_\odot)>11.2$ (see also \citealp{2012MNRAS.422.1714N}). The exact abundance of compact galaxies in the local Universe is still debated (e.g. \citealp{2010ApJ...712..226V}, \citealp{2012arXiv1211.1005P}, \citealp{2009ApJ...692L.118T}) and the impact of newly born galaxies is not fully understood (e.g Newman et al. 2012, \citealp{2012A&A...548A...7L}, Kaviraj et al. 2013). The morphological evolution of these massive galaxies and how it affects size evolution is also unclear (e.g. \citealp{2011arXiv1111.6993B}, \citealp{2012MNRAS.tmp..141H}, \citealp{2011ApJ...730...38V}, \citealp{2011ApJ...743L..15V}).

% Because of their different merger histories, hierarchical models predict that galaxy sizes at fixed redshift and stellar mass depend on environment \citep{2011arXiv1105.6043S}. In fact, the effect of mergers is enhanced in clusters and hence galaxies in larger haloes appear larger on average than galaxies in less massive haloes and the field \citep{2011arXiv1105.6043S}.

Environment is another powerful observable which can shed new light into the puzzle. In fact, some hierarchical models predict a significant environmental dependence in the sense that galaxies residing in more massive halos tend to be larger than galaxies in lower mass halos, at fixed stellar mass and redshift (e.g. \citealp{2011arXiv1105.6043S}). Unfortunately, observational studies at different redshifts have led to controversial results. Three works at $z\sim0$, $z<0.4$ and $z\sim1.2$  \citep{2009MNRAS.394.1213W, 2010MNRAS.402..282M, 2010ApJ...709..512R} did not find any significant trend of the mass--size relation with environment. \cite{2010ApJ...712..226V} found however a high fraction of \emph{super dense galaxies} in clusters in the local Universe, a result that is confirmed by Poggianti et al. (2012) who also claimed that early-type galaxies in clusters are slightly smaller than those living in the field at fixed stellar mass. At $z \approx 1.3,~$\cite{2012ApJ...745..130R} studied a sample of morphologically selected early-type galaxies in three different environments (field, cluster, groups) and found that, on average, for masses $10<log(M/M_\odot)<11.5$ cluster galaxies have either the same size or appear to be smaller at fixed stellar mass than field galaxies, depending on the stellar population model used.  More recently, \cite{2012MNRAS.tmp..141H}  did not detect any correlation with environment below $z\sim1$ up to the group scale ($log(M_h/M_\odot)<14$). On the other hand,  in the same stellar mass range but using a different definition for environment, \cite{2012MNRAS.419.3018C}  found exactly the opposite trend. Larger sizes in the cluster environment are also observed at $z=1.62$ by \cite{2012ApJ...750...93P} for passive galaxies with stellar masses larger than $log(M/M_\odot)\sim10.5$ and by \cite{2013arXiv1307.0003D} in a sample of clusters at $0.8<z<1.5$ with a similar selection (see also \citealp{2013arXiv1307.3247L}). 
The differences between these works are still to be understood and might come from different sample selections and/or the way environment is measured and/or low statistics at high redshift. 

%Another issue is how the position of the galaxy in a given halo (central vs. satellite) affects its subsequent evolution. Because of their particular position, merger models predict that galaxies residing at the centre (BCGs) of more massive haloes should, at fixed stellar mass, experience more mergers
%and thus be larger than their counterparts in less massive haloes
%, at least above $M_h>5 \times 10^{12} M_\odot$, according
%to the analysis of \citet{2011arXiv1105.6043S}. The latter, in fact, showed that BCGs should evolve
% much faster at $z<2$, and also end up being larger
%than other galaxies of similar mass. Current observational works, though, do not agree about the BCG size evolution and how BCG sizes compare to those of field and satellite galaxies. \cite{2009MNRAS.395.1491B} found that BCGs are larger than field and satellite galaxies at fixed stellar mass and that there is a steep evolution of their size from $z\sim0.3$ to present. The author also argues that minor dry mergers are the most probable mechanism to explain the build-up of these objects. In contrast, \cite{2009MNRAS.394.1213W} did not find a significant difference between the sizes of centrals and satellites in groups in the local Universe. 

%At higher z, \cite{2011ApJ...726...69A} find a significant size evolution of centrals from $z\sim0.6$ to present, but do not detect any evolution in their profile. This result is however at varaince with the work by \cite{2011MNRAS.414..445S} who studied a sample of high redshift BCGs and found a very mild evolution of the BCG size from $z\sim1$. 

In this paper, we revisit this issue by studying the mass-size relation of central and satellites ETGs in different environments selected from the Sloan Digital Sky Survey (SDSS) DR7 \citep{2009ApJS..182..543A} with an updated and accurate sample. The large statistics available make the SDSS the best sample to probe the environmental dependence of galaxy sizes. We probe an halo mass range  $12<log(M_h/M_\odot)<15$. 

%The SDSS observations are compared to predictions from the standard cosmological model, $\Lambda$CDM, obtained from the \cite{2010MNRAS.404.1111G} semi-analytical model and from the \cite{2011arXiv1105.6043S} model. 

%The letter is organized as follows. In section~\ref{sec:data} we describe the dataset and the sample selection and our main results are discussed in section~\ref{sec:results}. 

%Hierarchical models predict galaxies in larger haloes to be larger of $\approx 1.5$ times than galaxies in the field, because of their more extended merger histories. We find that the SDSS observations and the model predictions are at variance at $> 3 \sigma$. 

\section{ETG sample selection}
\label{sec:data}
%\subsection{Main sample}
%\label{sec:main}
We selected our ETG galaxy sample from the Sloan Digital Sky Survey DR7 spectroscopic sample \citep{2009ApJS..182..543A}. We select galaxies with an early-type morphology based on the galaxy morphological classification from Huertas--Company et al. (2011)\footnote{\url{http://gepicom04.obspm.fr/sdss\_morphology/Morphology\_2010.html}}. The authors performed a bayesian automated classification of the full SDSS DR7 spectroscopic sample based on support vector machines and associated to every galaxy a probability to be in four morphological classes (E, S0, Sab and Scd). In this work we select as ETGs those objects with a probability to be early-type (E or S0) greater than 0.8. Results do not change significantly if the probability threshold is changed between 0.5 and 0.8. 

To probe haloes of different mass, we use the group and cluster galaxy sample from \cite{2007ApJ...671..153Y}, updated to the DR7. \footnote{http://gax.shao.ac.cn/data/Group.html} This catalog of $\sim300,000$ clusters and groups ($\sim 30,000$ with more than 2 members) has been built using an automated halo-based group finder and provides an estimate of the halo mass in which galaxies live estimated through abundance matching. For this work, we restricted the analysis to groups with $z<0.09$ (for completeness reasons) and at least two members and also removed those objects affected by edge effects ($f_{edge}<0.6$). With this selection we expect that $\sim80\%$ of these groups have less than $\sim20\%$ contamination from interlopers \citep{2007ApJ...671..153Y}. We use as halo mass estimate, HM1, which is based on the characteristic luminosity of the group but results remain unchanged when using an halo mass estimate based on the characteristic stellar mass. The expected uncertainties on halo masses are 0.2-0.3 dex according to fig. 7 of \cite{2007ApJ...671..153Y} in which they compare estimated to true halo masses from a mock catalog. Also, the fact that the halo masses are derived through abundance matching using the luminosity (or stellar mass) might have an impact in our results. We will discuss the implications of these uncertainties in our main results.

Galaxy sizes are circularized effective radii obtained from the 2D Sersic
fit using the {\tt PyMorph} package \citep{2010MNRAS.409.1379V}, which can fit seeing convolved two components models
to observed surface brightness profiles.   The authors performed bulge to disk and single Sersic decompositions
to $\sim 7\times 10^5$ galaxies from the SDSS DR7. The algorithm is described
and tested in \cite{2012arXiv1211.6123M}. In particular, the sky estimate and how it affects size measurements is fully discussed in the mentioned work (see sec. 3.6). The authors showed through extensive simulations that the sky estimated with {\tt PyMorph} is underestimated by 0.1\% which has no major impact in the size estimate ($<10\%$). It is also shown that a bias in the sky value larger than $0.5\%$ is required to have a significant impact on the size ($>10\%$). For what concerns our work, the important result is therefore that sizes are unbiased with a typical scatter $<0.1$ dex (which depends on luminosity).For consistency with high redshift works we use here the sizes estimated from single Sersic fits which are shown to be less accurate than the ones obtained with two component models \citep{2012arXiv1211.6122B}. We have checked though that our results are unaffected by that choice. 

%Tests on synthetic images show
%that when the fitted functional form is the same as the one used to
%generate the image {\bf what happens if this is not the case?}, then {\tt PyMorph} returns accurate values of
%the measured parameters (e.g., total light, half-light radius, bulge-total
%ratio). .

Mass to light ratios have been obtained from the MPA-JHU DR7 release\footnote{\url{http://www.mpa-garching.mpg.de/SDSS/DR7}}. They are derived through SED fitting using BC03 synthesis population models \citep{2003MNRAS.344.1000B} and a Kroupa IMF following the procedure presented in \cite{2003MNRAS.341...33K} and Salim et al. (2007). We then convert to stellar masses by multiplying the $M/L$ of each galaxy by its luminosity estimated from the best fit Sersic model. In order to compare to models, we also apply a 0.05 dex shift to convert to a Chabrier IMF following \cite{2010MNRAS.404.2087B}. The typical error expected for photometrically derived stellar masses is $\sim0.2$ dex which is the value that will be used in the following (e.g. \citealp{2010MNRAS.404.2087B}).

The final sample contains $\sim 12,000$ ETGs with $log(M_*/M_\odot)>10.5$ and $z<0.09$, living in groups and clusters with halo masses from $M_h/M_\odot \sim10^{12.5}$ to $M_h/M_\odot \sim10^{15}$.

\section{Results}

\label{sec:results}

\subsection{Mass-size relation of ETGs in different environments}
\label{sec:mass_size}
\begin{figure*}
%%\epsscale{1.0}
%\resizebox{\hsize}{!}
{\includegraphics[scale=.6]{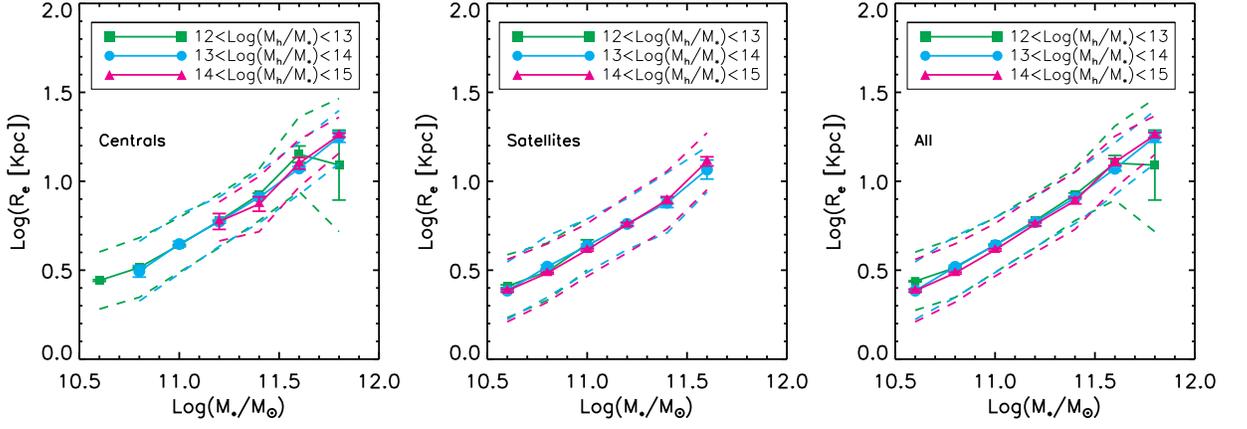}}
\caption{ Median stellar mass-size relation for centrals (left panel), satellites (middle panel) and all ETGs (right panel) living in different halo masses. Error bars are errors on the medians computed through bootstrapping and dashed lines show the $1-\sigma$ scatter.}
\label{fig:mass_size}
\end{figure*}

\begin{figure}
%%\epsscale{1.0}
\resizebox{\hsize}{!}{\includegraphics{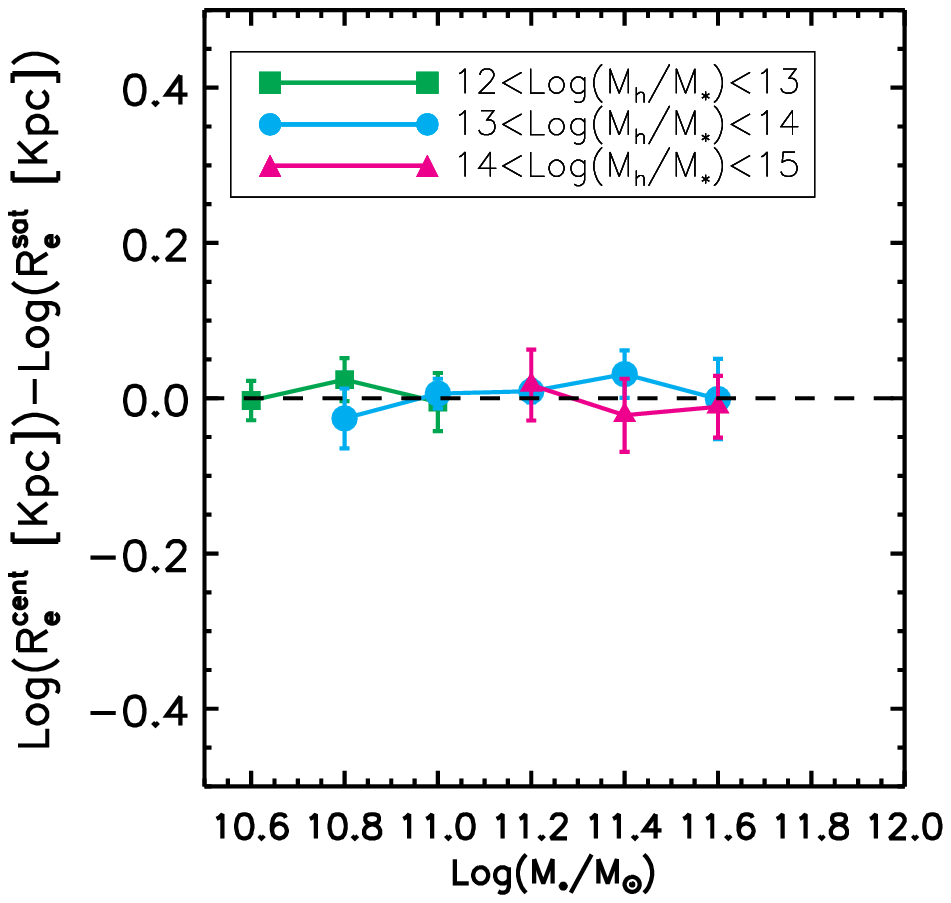}}
\caption{ Median size ratio of centrals and satellite galaxies for different halo masses as a function of stellar mass.}
\label{fig:cent_sat}
\end{figure}

Figure~\ref{fig:mass_size} shows the observational median stellar-mass relation for centrals, satellites and all galaxies living in haloes of increasing mass, typically corresponding to field, group and cluster environments. Central galaxies are defined in all this work as the most massive galaxies in a given halo. For some groups, the central is not the same galaxy defined by Yang et al. (2007) because we reprocessed stellar masses using the Sersic luminosity as explained in section~\ref{sec:data}. The main results remain however unchanged when using the original definition. Our first result is that the mass-size relation of satellites and central galaxies do not show any significant trend with environment, i.e they present similar mass--size relations independently of the mass of the host. We confirm the preliminary results by \cite{2009MNRAS.394.1213W} but with a much larger sample and better defined sizes and morphological classification. We notice however than in a recent work using an independent dataset, \cite{2012arXiv1211.1005P} found that the mass-size relation of cluster galaxies lies slightly below ($\sim1\sigma$) the relation for field galaxies (see also \citealp{2010ApJ...712..226V}). It is still unclear what can make the difference and it certainly requires further investigation. Morphological selection could for example play a role since  \cite{2012arXiv1211.1005P} sample seems to be dominated by lenticular galaxies which have been shown to be systematically smaller than elliptical galaxies at fixed stellar mass (\citealp{2012arXiv1211.6122B,2012MNRAS.tmp..141H} see also section~\ref{sec:sel-effects}) when the size is estimated with a single sersic profile and then circularized. In fact, in the \cite{2012arXiv1211.1005P} sample, there are $\sim50\%$ S0s in clusters while only $\sim30\%$ in the field (private communication) which would partially explain the fact that they find smaller galaxies in clusters. 
In any case what seems to arise from these works is that if there is a difference with environment at $z\sim0$ it must be small.
Moreover, at fixed halo-mass, satellites and centrals present similar mass-size normalizations (fig.~\ref{fig:cent_sat}) and scatters which suggests that the mass-size relation is universal, independently of the position of the galaxy in the halo. 
%In the following, we will focus on the high mass end and try to properly quantify this apparent lack of dependence taking into account our observational errors. 

%The mass-size relation of central galaxies depends on the large-scale environment, specially for galaxies above $\sim10^{11}$ solar masses. The general trend is that, at fixed stellar mass centrals living in massive haloes tend to be larger than the same galaxies living in less massive haloes. The difference increases with increasing stellar mass.  Therefore, central galaxies of $10^{11.5}$ solar masses living in $\sim10^{13}M_\odot$ haloes are 3 times smaller than galaxies in the center of  $\sim10^{14}M_\odot$ haloes, while the difference is size is reduced to a factor of $\sim1.3$ for galaxies of $\sim10^{11}$ solar masses. 
%In contrast satellite galaxies do not show this trend, i.e. they present similar mass--size relations independently of the mass of the host. Interestingly, the trend is still visible in the whole population simply because above $\sim10^{11.5}$ solar masses, the population of ETGs is completely dominated by central galaxies. 

%The differences between central and satellites at high stellar masses can be interpreted as a signature of mergers.  

\subsection{$M_h-\gamma$ relation of massive ETGs}
\label{sec:mh-gamma}
%Hierarchical models predict indeed that the assembly of the most massive ETGs ($M_*/M_\odot>10^{11}$) is mainly driven by mergers. Since galaxies in bigger haloes experienced on average
%more mergers during their lifetime (especially centrals), if mergers increase galaxy sizes we should expect this population to exhibit an environmental effect pointing towards larger galaxies living in larger haloes. 

The fact that galaxies of similar mass share similar size distributions irrespective of their environment, does not directly rule out some intrinsic environmental dependence. In fact, the intrinsic scatter of the mass-size relation for massive ETGs ($\sim0.2$ dex - e.g. \citealp{2011MNRAS.412L...6B, 2012arXiv1211.6122B}, see also fig.~\ref{fig:mass_size}) puts an upper limit to that effect, i.e. galaxies in massive haloes can be at most a factor 3 ($2\times 10^{0.2}$) larger than the same galaxies living in small haloes. As a result, the detection of the signal might be difficult given the observational uncertainties in the different  variables at play (sizes, halo masses, galaxy classification, stellar masses) which can reduce any observed trend (see sections~\ref{sec:errors}). 

Therefore, in the next two sections we want to focus on the high mass end of the mass function (where the impact of mergers should be more pronounced) and look in detail for environmental effects taking into account as much as possible the effects of observational biases and errors.

\begin{figure*}
%%\epsscale{1.0}
%\resizebox{\hsize}{!}
{\includegraphics[scale=0.6]{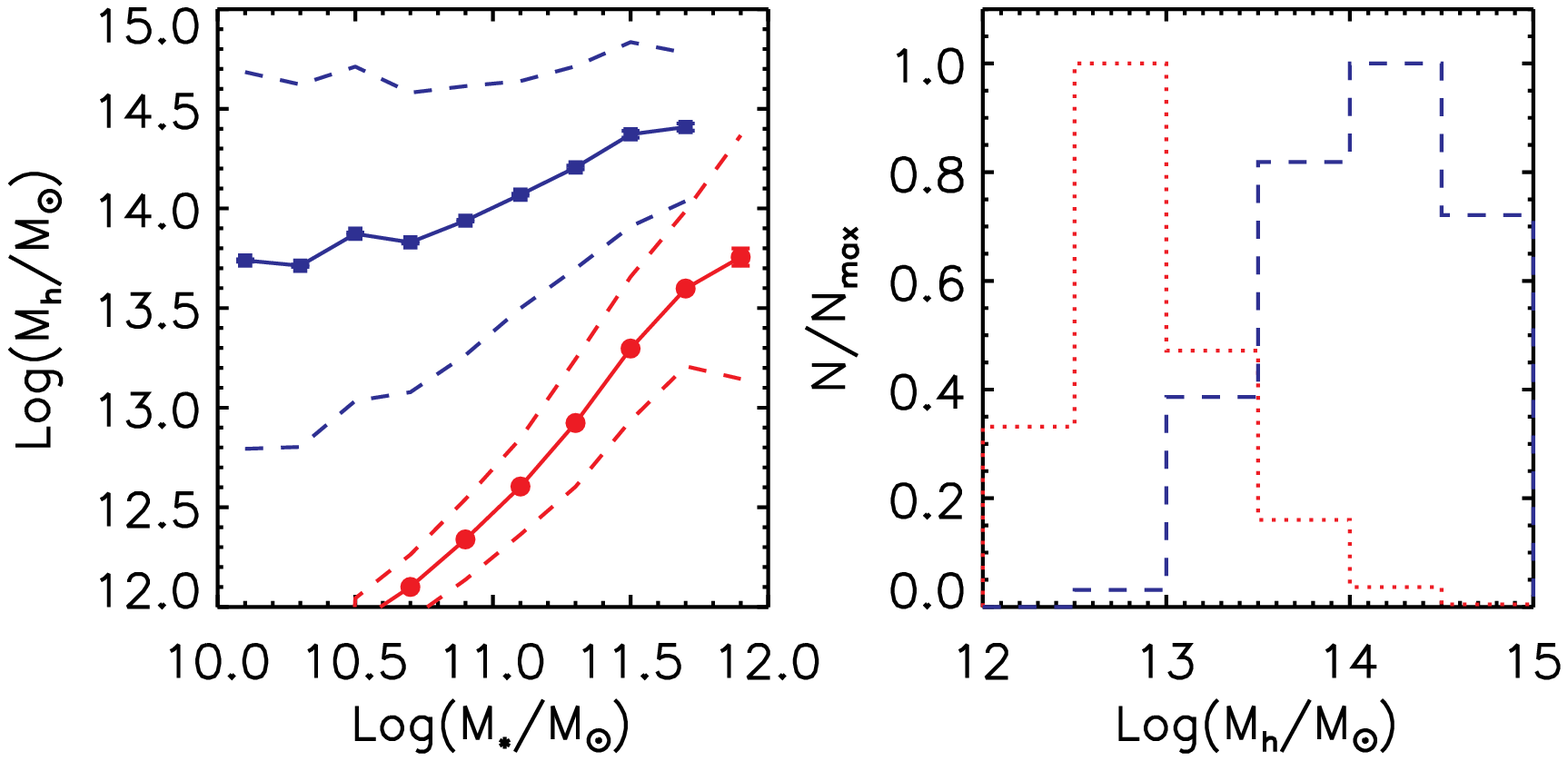}}
\caption{ Left: Median $M_*-M_h$ relation for central ETGs (solid red, filled circles) and satellites ETGs (solid blue, filled squares). Dashed lines show the $1-\sigma$ errors. Right: Histogram of halo masses for ETGs more massive than $10^{11}$. Dotted red: central galaxies, dashed blue: satellite galaxies.}
\label{fig:mhalo_mstar}
\end{figure*}

To this purpose we analyze the $M_h-R_e$ relation, which gives the median size of ETGs at fixed stellar mass as a function of environment. While there is a well known correlation between the mass of the halo and the stellar mass of galaxies populating it (e.g. \citealp{2004ApJ...617..879L}), the scatter of that relation is large enough so that galaxies of a fixed stellar mass populate a large range of haloes (fig.~\ref{fig:mhalo_mstar}), allowing a study of environmental effects at fixed stellar mass. 

\begin{figure}
\epsscale{0.3}
%\resizebox{\vsize}{!}
{\includegraphics[scale=0.6]{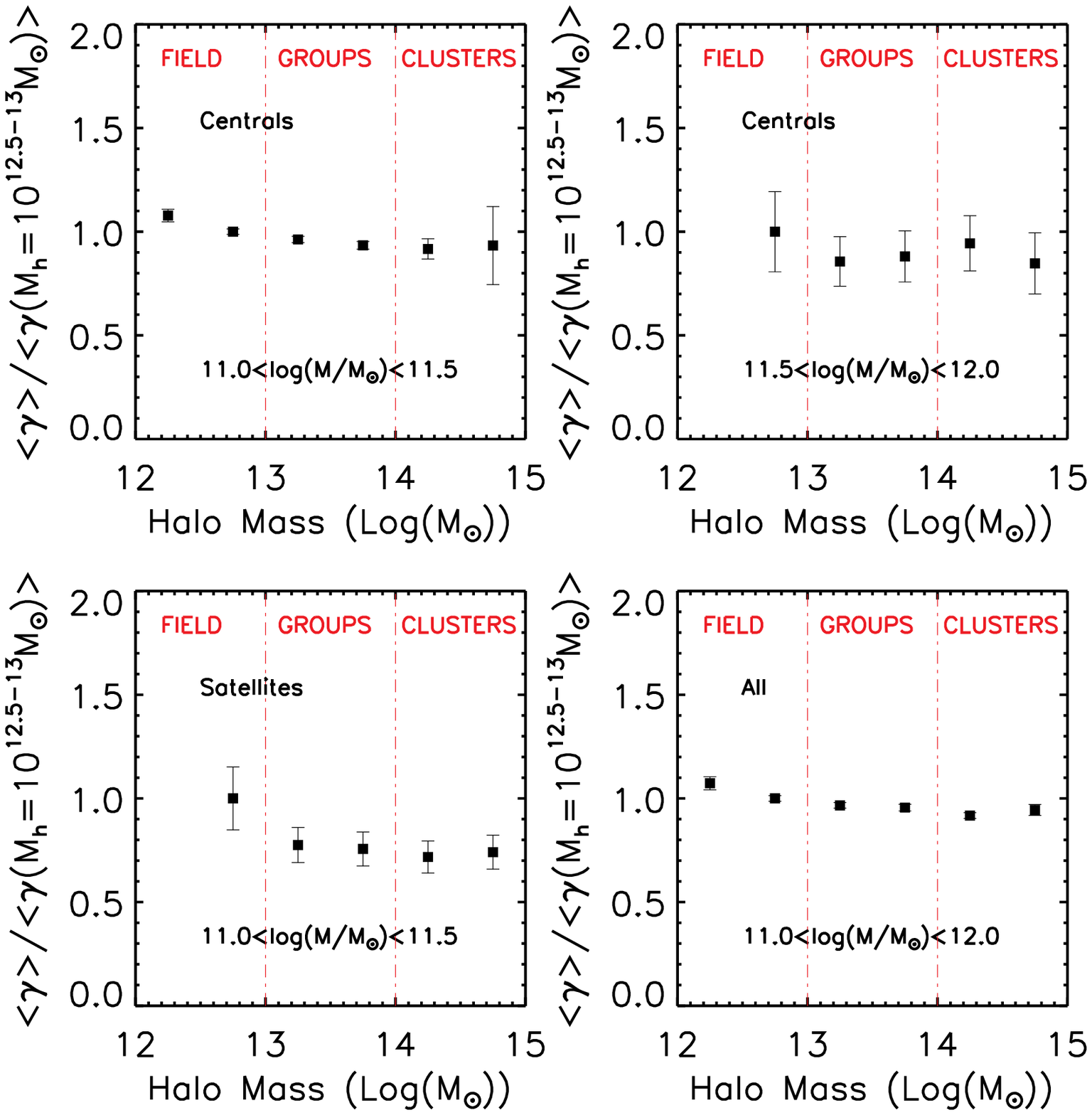}}
\caption{Size of central (top panel), satellites (bottom left panel)  and all ETGs (bottom right panel) as a function of halo mass in the SDSS in different stellar mass bins as labelled. Values have been normalized so that, by definition, the field observed value at an halo mass of $log(M_h/M_\odot)=12.5-13$ is equal to 1. Errors are errors on the median values computed through bootstrapping (see text for details).}
\label{fig:re_halo_yang}
\end{figure}

Our main results are shown in figure~\ref{fig:re_halo_yang} for central galaxies in two stellar mass bins ( $11<log(M_*/M_\odot)<11.5$ and $11.5<log(M_*/M_\odot)<12$) and for satellite galaxies in one single stellar mass bin (very massive satellites only exist in massive haloes). We use stellar mass bins relatively large to increase statistics and minimize the impact of errors in stellar mass ($\sim0.2$ dex). However, this choice could induce spurious correlations between size and environment, since more massive galaxies and hence larger, preferentially live in massive haloes given the existing correlation between $M_h$ and $M_*$. In order to get rid of this effect, we use normalized sizes ($\gamma$) following a similar procedure than the one explained in Newman et al. (2011) and Cimatti et al. (2012):

\begin{equation}
Log_{10}(\gamma)=Log_{10} (Re)+\beta \left (11-Log_{10}(M_*\right) )
\label{eq:gamma}
\end{equation}
where, $\beta$ is the slope of the $M_*-R_e$ relation in the considered mass interval and  $R_e$ is the effective radius. We use here a value of $\beta=0.83$ which is the slope measured in the mass-size relation for galaxies witj $log(M_*/M_\odot)>11$. 
%We better quantify these differences and compare our results with predictions of semi-analytic hierarchical models in figures~\ref{fig:re_halo_yang} to \ref{fig:re_halo_yang_sat} in which we plot the halo mass - size relation for centrals, satellites and all galaxies respectively in different stellar mass bins going from $10^{11}M_\odot$ to $10^{12}M\odot$. For each halo mass and stellar mass bin we compute the median size of the galaxies living in the given environment. There is obviously a correlation between stellar mass and halo mass, as seen in figure~\ref{fig:mass_size}, i.e. very massive galaxies tend to be located in massive haloes. However the scatter is large enough so that we can look at the population of galaxies living in different haloes at fixed stellar mass in particular above $10^{11}M\odot$. We also normalize sizes in figures~\ref{fig:re_halo_yang} to \ref{fig:re_halo_yang_sat} have been mass-normalized following Newman et al. (2011) and Cimatii et al. (2012) ( $\gamma=R_e\times (10^{11}M_\odot/M_*)^{0.57}$) to reduce the effect of the correlation of stellar mass and halo mass.

 Finally, since we are interested in relative differences between the different environments, we normalize all sizes to the median size in the halo mass bin $M_h/M_\odot=10^{12.5}-10^{13}$. That way, by definition, all median sizes in that halo are equal to one. Uncertainties on the median values are computed through bootstrapping, i.e. we repeat the computation of each value $1,000$ times removing one element each time and compute the error as the scatter error of all the measurements. 

The most striking result is that the $\gamma-M_h$ relation is essentially flat independently of the stellar mass and of the position of the galaxy in the halo, i.e. sizes of massive ETGs are the same at all environments within the errors. In the next section we discuss the robustness of this result to observational errors and selection effects.

\section{Discussion}
\label{sec:discussion}

%Only very massive centrals do show a kind of trend if we consider those galaxies living in the less massive haloes ($M_h=10^{12-12.5}$), however there are very few objects in that bin and the uncertainty is large {\bf I would give a table with number of galaxies for each bin}. 

%However, other mechanisms present in galaxy evolution models such as strong disk instabilities may play a non-negligible role. Detailed discussion of theoretical models will be presented in separate work (Shankar et al. in preparation).

%When considering satellites however, both models and observations show a lack of dependence of the galaxy-size with environment. This is expected since satellite-satellite mergers are very rare in a hierarchical scenario.  

\subsection{Selection effects}
\label{sec:sel-effects}
In the previous sections, we have shown results for galaxies selected at fixed stellar mass and with early-type morphology. We investigate in this section the impact of the selections in our results. 

\subsubsection{Stellar mass selection}
\begin{figure}
\epsscale{0.3}
%\resizebox{\vsize}{!}
{\includegraphics[scale=0.6]{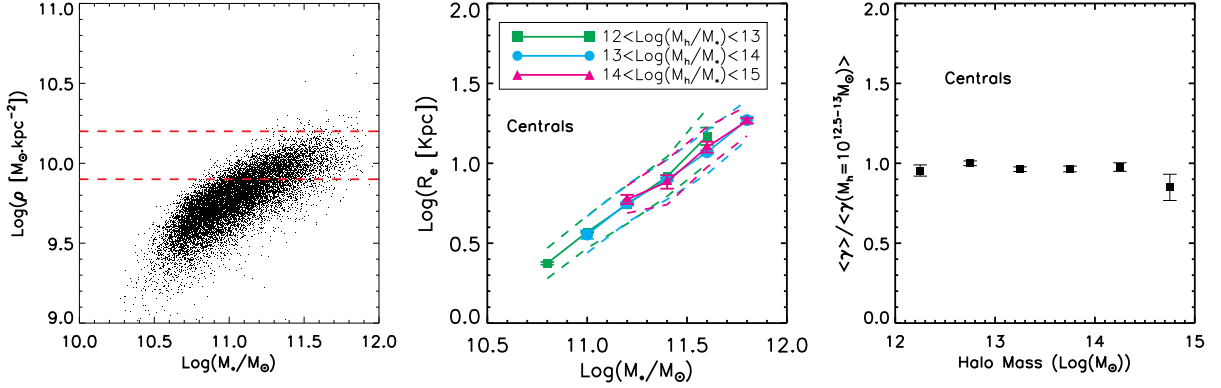}}
\caption{$M_h-\gamma$ plane at fixed central mass density. Left panel shows the relation between stellar mass and central density for central ETGs. Red lines indicate the range of central mass densities considered to produce the plots on the middle and right panels.The middle panel shows the mass-size relation for the selected galaxies in three environments. The right panel shows the $M_h-\gamma$ plane for ETGs with $10^{9.8}<\rho<10^{10.1}$. The trend is still consistent with flat. Values have been normalized so that, by definition, the field observed value at an halo mass of $log(M_h/M_\odot)=12.5-13$ is equal to 1. Errors in models and observations are errors on the median values computed through bootstrapping.}
\label{fig:re_halo_rho}
\end{figure}

While in the expansion scenario (see sec.~\ref{sec:intro}) galaxies puff-up at constant stellar mass, in the merger model galaxies contemporarily also grow in mass by a factor 2-3 \citep{2009ApJ...699L.178N}. Studying environmental dependence at fixed stellar mass may thus not be the ideal choice to test hierarchical scenarios even though we expect this fact to have a small effect in our results given the relatively large bins of stellar mass used (0.5 dex). What seems clear is that, to be effective in increasing sizes, minor mergers should preferentially increase the outskirts of the stellar distributions leaving the central regions more or less intact. Therefore, one alternative way to probe environmental effects could be to fix central mass density instead of total stellar mass density. Results are shown in figure~\ref{fig:re_halo_rho} for central galaxies with central densities that roughly correspond to galaxies of $\sim10^{11}$ solar masses. Projected central densities are computed in the inner 1 Kpc, using the best fit profiles as done for instance by \cite{2012MNRAS.422.3107S}. The observed trend in the $M_h-\gamma$ plane is still consistent with flat, confirming our previous results. We notice that the $\beta$ normalization factor to compute $\gamma$ (see eq.~\ref{eq:gamma}) is larger than the one used at fixed stellar mass since the mass-size relation is steeper when the central mass density is fixed (see middle panel of figure~\ref{fig:re_halo_rho}).

%By looking at the at the $M_h-\gamma$ plane at fixed stellar mass we are intentionally neglecting the fact that a merger event increases the stellar mass of the final object. In fact, given the relatively large bins of stellar mass used (0.5 dex), we expect . One alternative to check this assumption is to look at the effect of environment at fixed central mass density. In fact, since minor mergers are thought to affect mainly the outskirts of the galaxy, at first level, the central parts stay unaffected. 

%There is a hint of an increasing  trend which is not present in fig.~\ref{fig:re_halo_yang} but we believe it is a consequence of the mass-size relation since the bin in stellar mass is larger and therefore the normalization of sizes does not fully remove the correlation. Notice also that since Sersic indices are not an output of the models we do not over plot model predictions in this figure. 

\subsubsection{Morphology selection}
\begin{figure}
%%\epsscale{1.0}
%\resizebox{\hsize}{!}
{\includegraphics[scale=0.6]{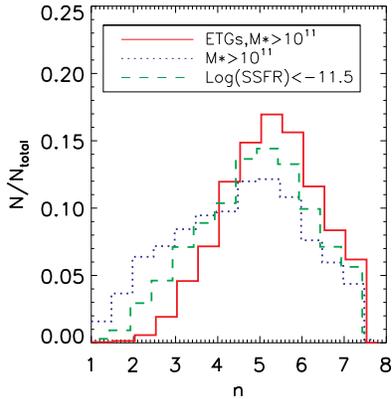}}
\caption{ Sersic index distribution for galaxies with $Log(M_*/M_\odot)>11$ for different selections. }
\label{fig:sersic}
\end{figure}

\begin{figure}
\epsscale{1.0}
%\resizebox{\vsize}{!}
{\includegraphics[scale=1.0]{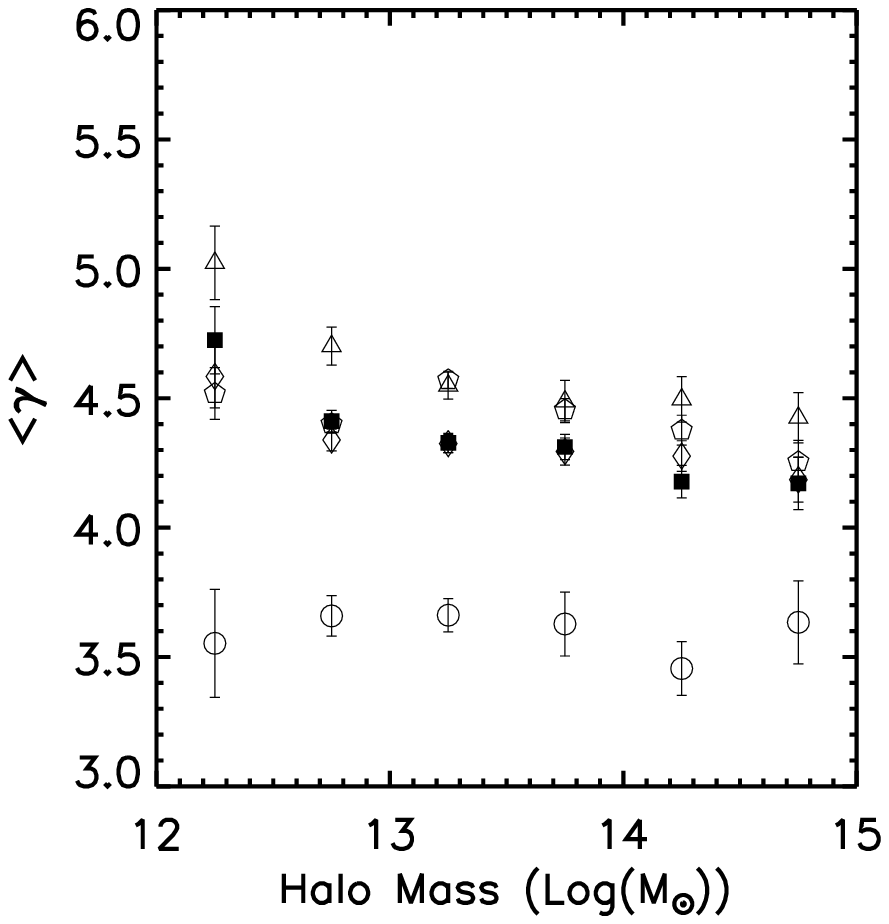}}
\caption{$M_h-\gamma$ for all $10^{11}<M_*<10^{12}$ galaxies (empty pentagons), all passive (empty diamonds), ellipticals (empty triangles), lenticulars (empty circles) and ETGs (filled squares). All selections give results consistent with flat. S0s are systematically smaller at all environments when circularized sizes are used.}
\label{fig:gamma_selections}
\end{figure}

An observational signature of merger models (see sec.~\ref{sec:intro}) should be
       a systematic increase in the Sersic index with time, 
       while the expansion scenario tends to preserve the original profile, at least up to 50\% of mass loss 
       (Ragone \& Granato 2011). Thus, in a hierarchical scenario, more evolved systems (i.e., with more mergers) are expected to have, on average, higher Sersic indices (e.g., Hopkins et al. 2009). %However, it is also known empirically 
       %that Sersic index broadly increases with stellar mass (REF.).
        By selecting only early-type galaxies in our study, we thus might be biased 
       towards higher values of the Sersic index, and not being properly considering an enough wide dynamic range to probe different growth 
       histories. In other words, our selection of ETGs might artiÞcially flatten the signal since we might
       be preferentially selecting objects with high Sersic index, with an assembly history 
       possibly dominated by mergers, and missing objects with lower Sersic index mostly grown via insitu star formation (which could be more common in low density environments). On the other hand, if we do not apply any morphological selection, we might detect an environmental signal if the morphological mixing changes significantly with environment since it is well known that size and morphologies correlate.http://www.autourdebebe.com/catalogue/promotions.aspx?idPromo=4718593
       In figure~\ref{fig:sersic} we first show that the Sersic index distribution of the selected galaxies is broad (even if dominated by high values) indicating that we are indeed probing different formation histories. The distributions change slightly if the selection is based on stellar mass or star formation (instead of morphology) as expected but most importantly, our main results discussed previously remain unchanged. In figure~\ref{fig:gamma_selections} we show indeed the $\gamma-M_h$ plane for different selections in the stellar mass range $11<log(M_*/M_\odot)<11.5$ (ETGs, lenticulars, ellipticals, passive or no selection at all). The selection of passive galaxies is based on the total median specific star formation rates (SSFRs) computed by \cite{2004MNRAS.351.1151B}\footnote{http://www.mpa-garching.mpg.de/SDSS/DR7/sfrs.html}. More precisely we select as passive galaxies those objects with $-15<Log(SSFR (yr^{-1}))<-11.5$ based on the bimodal distribution of the SSFR. We do not normalize here to explicitly measure the different normalizations between the different selections.  All selections show a behavior consistent with flat. The samples without morphological selection present slightly higher sizes due to the contamination of spiral galaxies. The most noticeable difference appears in the lenticular population which is systematically $\sim15\%$ smaller than all the other selections (including ellipticals). This fact has already been noticed by Bernardi et al. (2012) and Huertas-Company et al. (2013) at $z\sim1$ and it is consistent with the recent claims that the most compact galaxies have a disk component (e.g. \citealp{2012ApJ...751...45T}, \citealp{2011ApJ...730...38V}, \citealp{2013ApJ...762...83C}). We notice that most of this effect is due to the way sizes are computed, i.e. we use a single Sersic to model the light of a galaxy with two components by definition and the radii are circularized with values of $b/a$ which are smaller on average since S0s are better identified when they present high inclinations.

%Same trends already observed in figure~\ref{fig:mass_size} are recognized in these figures. The halo mass - size relation for centrals becomes steeper at increasing stellar masses as seen in figure~\ref{fig:re_halo_yang_cent}. The trend is very well reproduced by hierarchical models and it is a direct consequence of mergers simply because galaxies lying in the centers of massive haloes experience more mergers than the ones in less massive haloes.

%begin{figure}[h!]
%%\epsscale{1.0}
%\resizebox{\hsize}{!}{\includegraphics{re_halo_yang_MA_betaCHange_sat_norm2.ps}}
%\caption{Same as figure~\ref{fig:re_halo_yang} but for satellite galaxies.}
%\label{fig:re_halo_yang_sat}
%\end{figure}

\subsection{Can errors wash out the signal?}
\label{sec:errors}

As previously stated, the scatter of the mass-size relation for massive galaxies is not very large, a factor 1.5-2 typically. Therefore, the environmental signature is bounded to a factor 3 to 4 at most. It is important then to properly understand if the lack of dependence on environment we measure is a consequence of observational errors in the different parameters involved ($M_h$,$M_*$,$R_e$), which could wash out the signal or a real signature. 

We investigate in this section through Monte Carlo simulations what is the global effect of statistical errors on a possible existing signal. To that purpose, we create an artificial trend with environment within the constraints imposed by the scatter of the real mass-size relation i.e at fixed stellar mass, ETGs living in low mass haloes can at most be $\sim3$ times smaller than their counterparts living in the most massive haloes (twice the scatter of the mass-size relation).Therefore, to each galaxy, given its measured stellar mass, size and halo mass from the real data, we add a positive shift to its size which scales linearly with its halo mass: 

\begin{equation}
%M_h(M_*,R_e)=\overline{M_h(M_*)}+\sigma_{M_h}(M_*)\frac{R_e-\overline{R_e(M_*)}}{\sigma_{R_e}(M_*)}
R_e^{sim}=R_e+(Log(M_h)-12)\times \kappa
\label{eq:simus}
\end{equation}

Increasing values of $\kappa$ will produce larger environmental effects. This way, smaller galaxies at fixed stellar mass will preferentially be associated to smaller halos.  We then investigate the effect of adding increasing gaussian random errors to stellar mass, size and halo mass. Notice, that with this procedure we are assuming that the observed mass-size relation is the intrinsic one (before convolution with errors) which is an approximation since it already contains errors. Nevertheless, the conclusions drawn from our Monte Carlo experiments are independent of the exact choice of initial condition, as long as the all the variables are properly updated after convolution with errors.

The way errors correlate on the three variables is not a trivial question and clearly depends on the dataset and the way the three parameters are derived from an observational point of view. We have explored two extreme cases in this work which fully bracket the whole range of meaningful possibilities. First, we have trivially considered a scenario in which errors in the three variables are completely independent. Then we worked out a second possibility characterized by the three variables being instead fully correlated, following the constraints imposed by how our sample of (central and satellites) galaxies was built (see below).

\begin{figure*}
%%\epsscale{1.0}
{\includegraphics[scale=0.65]{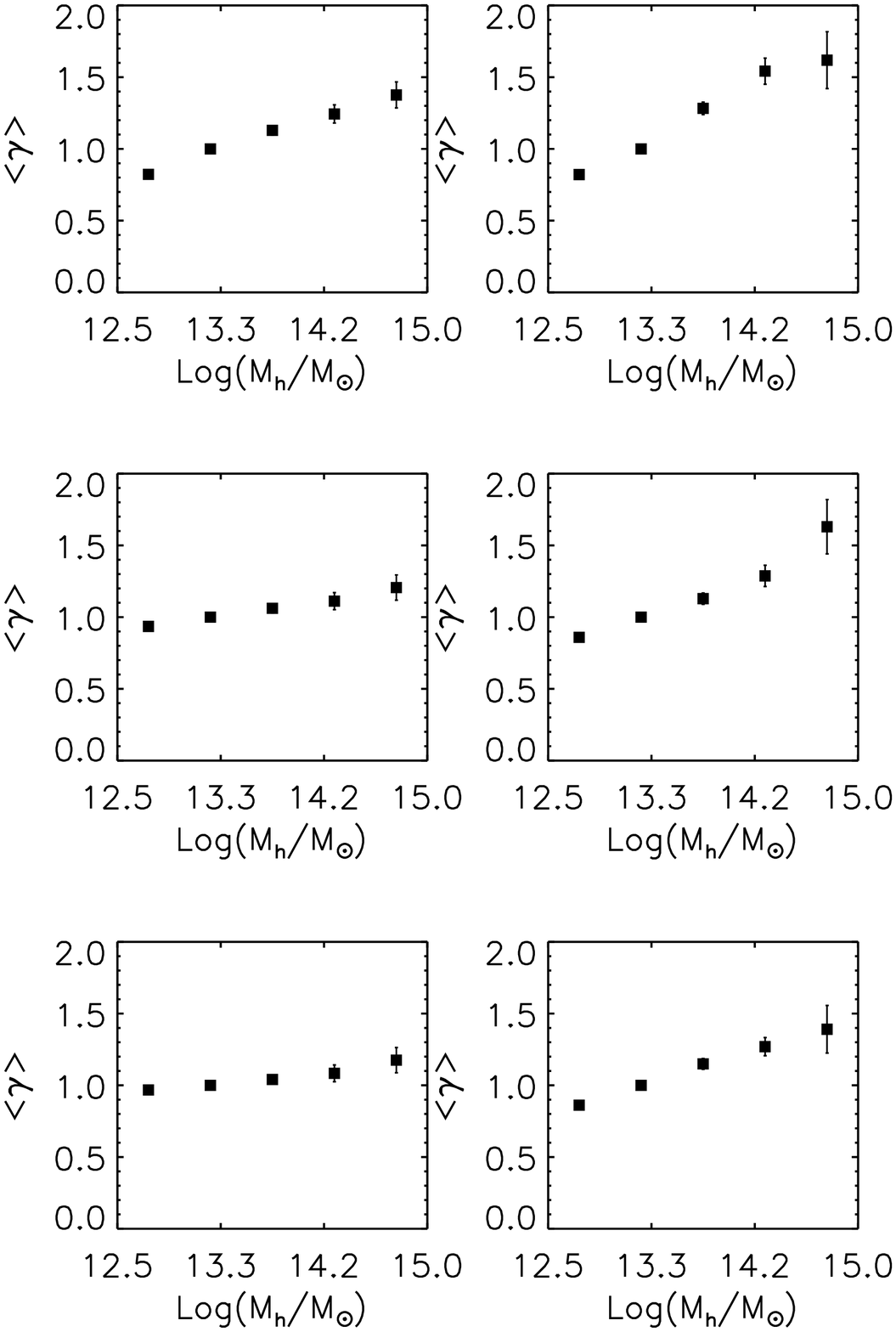}}
\caption{ Results of Montecarlo tests to assess the sensitivity to observational errors of an eventual environmental effect on the sizes of central ETGs when errors are uncorrelated. The left column shows different simulated $M_h-\gamma$ relations without errors  with decreasing values of $\kappa$, i.e. $4.0$, $2.0$ and $1.5$ from top to bottom (see text for details). The right column shows the resulting $M_h-\gamma$ plane after adding expected uncertainties on halo masses, stellar masses and sizes independently. The signal is enhanced under this hypothesis.}
\label{fig:mc_tests_un_cen}
\end{figure*}

Our first finding is that, if errors are uncorrelated, the signal with environment is basically preserved independently of the errors or even enhanced in the case of central galaxies (fig.~\ref{fig:mc_tests_un_cen}). The main reason behind this behavior is that halos are not populated in a uniform way by central galaxies in terms of stellar mass (fig. 3) in the sense that below $10^{11}$ solar masses, central galaxies preferentially populate small halos.  Adding errors to stellar mass (without updating the other variables) will then systematically add a population of new small galaxies living in small halos which will maintain or enhance an environmental signal.  

Since the independence assumption does not seem to be very realistic, at least for our sample, in which halo masses are determined using abundance matching with the group stellar mass function (Yang et al. 2007), we will below preferentially focus on scenario with correlated errors.

In fact, stellar masses and sizes are also correlated since the choice of a given model light profile with respect to another to fit the image will translate into an error in the total luminosity which is converted in to an error in stellar mass. \cite{2013arXiv1304.7778B} estimate that the maximum systematic error on magnitude can be for luminous galaxies of the order of 0.5 mag (see their fig.~1), which translates into 0.2 dex in luminosity, thus a 0.2 dex in stellar mass. On the other hand the same systematic shift produces up to 0.2 dex in size, thus a correlation of the type $\Delta Log Re\sim \Delta Log M_*$. This is the maximum correlation possible reported in the literature so far between $Log M_*$ and $Log R_e$. This systematic error is at least one order of magnitude larger than the statistical error on size (Meert et al. 2013), which we neglect.

Concerning halo masses, Yang et al. (2007) use abundance matching between the total stellar mass of the group/cluster and the halo mass to assign FoF halo masses to galaxies. Thus halo mass and galaxy mass are fully correlated by the cumulative relation between stellar mass function and halo mass function. When we assign gaussian errors in our simulations, in practice we convolve the stellar mass function with a gaussian thus increasing the number density of massive galaxies. Therefore this changes the mapping between stellar mass and halo mass. To properly take this effect into account we use the Millennium simulation and the Guo et al. model (2010). We compute in fact the stellar mass function of centrals with and without statistical and systematic errors as described in the previous paragraph, and each time compute the median $M_*-M_h$ relation to quantify the median shift in the halo mass at fixed stellar mass. Clearly both the stellar mass function and cosmology used in Yang et al. (2007) will be a little different, but we do not expect these changes to make any major impact in our conclusions given that we are mainly interested in the median shift not in the absolute value of $M_h$.

Results are shown in figures~\ref{fig:mc_tests_corr_cen} and~\ref{fig:mc_tests_corr_all} for decreasing values of $\kappa$ for centrals and for all galaxies respectively. The left column shows the signal before adding errors and the right column the same signal once errors are incorporated in the way just described. When correlated errors are included, the environmental signal tends to be washed out. Thus, a size ratio between cluster and field galaxies of $\sim1.4$ or larger will be detected at more than $3 \sigma$ even even after maximizing the potential effect of correlated errors. Size ratios lower than $1.4$ would be detected with small significance or not detected at all. 

These simulations show that despite of the different systematics involved in the determination of stellar masses, sizes and specially halo masses in the sample used in this work, we should be able to detect any size difference greater than a factor of $1.4$ between galaxies residing in extreme environments.

 \begin{figure*}
%%\epsscale{1.0}
{\includegraphics[scale=0.65]{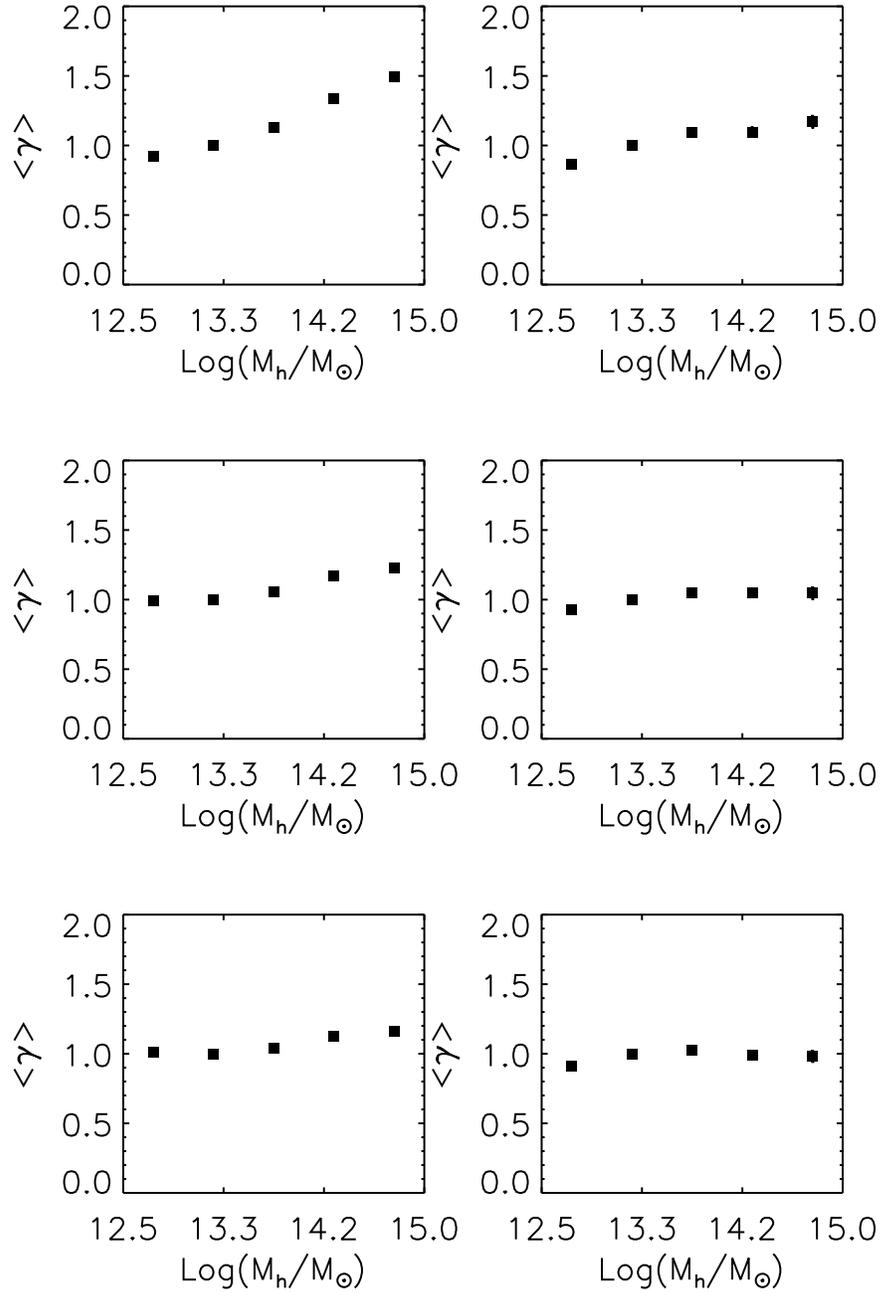}}
\caption{ Same as previous figure for all galaxies (centrals and satellites). When all galaxies are considered, any environmental dependence lower than $30\%$ would remain undetected if errors are added in an independent way.}
\label{fig:mc_tests_un_all}
\end{figure*}

 \begin{figure*}
%%\epsscale{1.0}
{\includegraphics[scale=0.65]{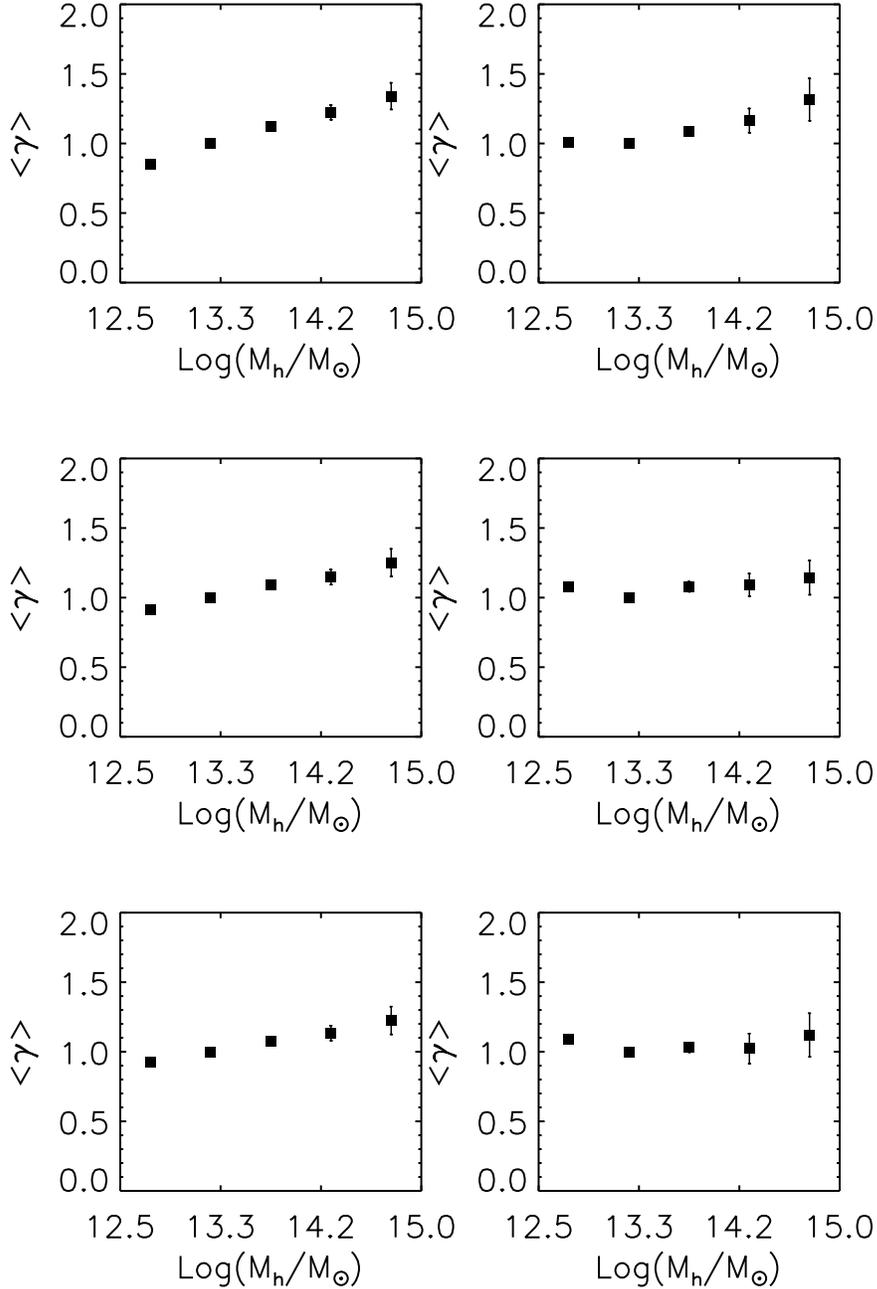}}
\caption{ Results of Montecarlo tests to assess the sensitivity to observational errors of an eventual environmental effect on the sizes of central ETGs when errors are correlated. The left column shows different simulated $M_h-\gamma$ relations without errors (see text for details). The right column shows the resulting $M_h-\gamma$ plane after adding expected uncertainties on halo masses, stellar masses and sizes as explained in the text. An environmental effect larger than $30\%$ should be detected at $3\sigma$ level given the errors in our sample.}
\label{fig:mc_tests_corr_cen}
\end{figure*}

 \begin{figure*}
%%\epsscale{1.0}
{\includegraphics[scale=0.65]{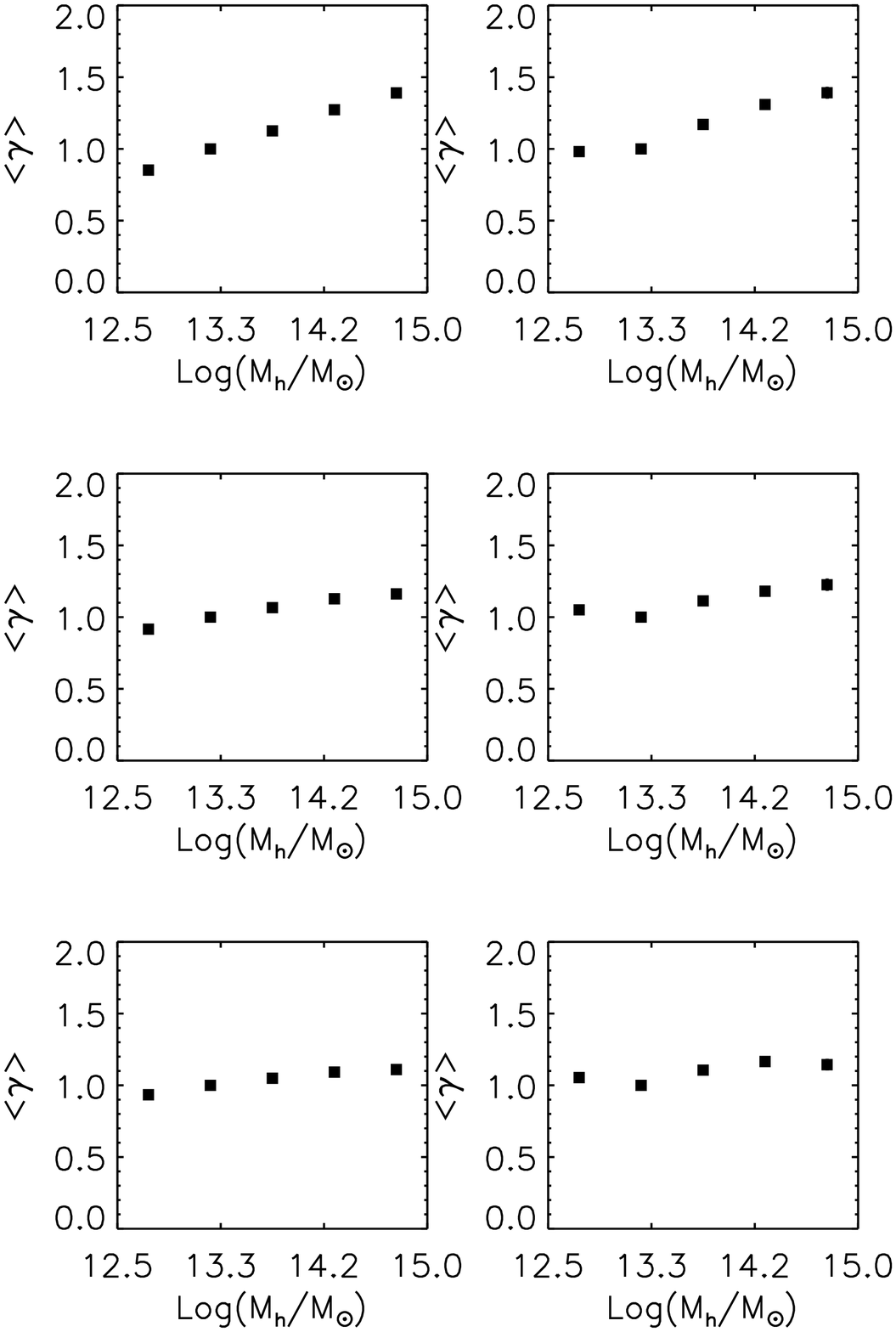}}
\caption{ Same as previous figure for all galaxies (centrals and satellites). }
\label{fig:mc_tests_corr_all}
\end{figure*}

\subsection{Effects of interlopers}

\begin{figure}
\epsscale{0.3}
%\resizebox{\vsize}{!}
{\includegraphics[scale=0.6]{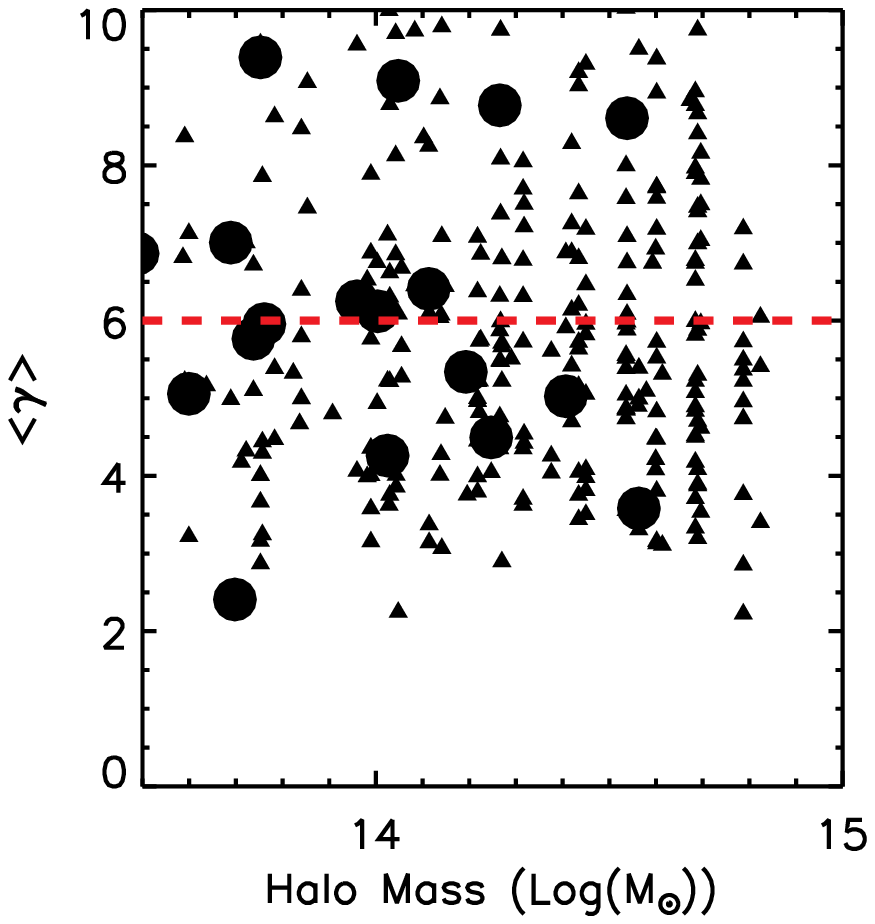}}
\caption{Mass-normalized size ($\gamma$) of central (big circles) and satellites (triangles) ETGs as a function of halo mass in the cluster sample of \cite{2007A&A...471...17A}. No correlation between size and halo mass is detected at the cluster scale (correlation coefficient is $\sim0.1$). The red dashed line shows the median $\gamma$ value measured in the SDSS field sample extrapolated to the cluster scale in order to check that there is no any significant trend.}
\label{fig:clusters_alfonso}
\end{figure}

As an additional check to estimate the effects of errors in halo mass and interlopers in the membership selection, we defined a smaller but robust control sample of spectroscopically confirmed members of massive clusters based on the selection of \cite{2007A&A...471...17A}. The sample is made of  88 clusters with known redshift at $z<0.1$ from the catalogues of \cite{1989ApJS...70....1A}, \cite{1961cgcg.book.....Z}, \cite{2000ApJS..129..435B} and \cite{1999A&A...349..389V}~that have been mapped by the SDSS-DR4 \citep{2000AJ....120.1579Y}. Cluster membership has been obtained using the velocity information from SDSS-DR4 by a combination of two algorithms. In a first step the ZHG algorithm was applied. In a second step, the cluster membership was refined by the applications of the KMM algorithm. The final sample contains a total of 10865 galaxies as cluster members (see \citealp{2007A&A...471...17A} for more details). Halo masses of those clusters were estimated independently based on numerical N-body simulations using Eq (2) of \cite{2006A&A...456...23B} rescaled for cluster redshift and cosmology. The errors in the mass estimations were obtained by propagating the errors in this equation. Sizes, stellar masses and morphologies come from the same catalogs than for the main sample (see sec.~\ref{sec:data}). We still find similar results using this independent sample. The Pearson correlation coefficient between halo mass and size is $\sim0.1$ proving that there is no correlation at the cluster scale and also the values of $\gamma$ are consistent with the ones measured from the field SDSS sample (fig.~\ref{fig:clusters_alfonso}). 

\section{Conclusions}

We have analyzed a sample of $\sim 12, 000$ local ETGs from the SDSS DR7, selected in different environments. Our main results are the following:

\begin{itemize}

\item The mass--size relation of ETGs in the local Universe does not significantly depend on environment. At fixed stellar mass (or fixed central stellar mass density), galaxies residing in clusters  have similar sizes than the ones living in the field. 

\item The mass-size relation does not depend either on the position of the galaxy in the halo, satellites and central galaxies follow similar mass-size relations.

\item If we focus on the high-mass end of the galaxy population ($log(M_*/M_\odot)>11$), we have shown that, given our estimated observational uncertainties and the size of our sample, we can rule out any size difference between cluster and field ETGs larger than 30-40\% at $3\sigma$. The flatness of the $R_e-M_h$ relation is therefore an intrinsic property and not a consequence of observational uncertainties.

\item The result is also robust to different galaxy selections. If galaxies are selected based on morphology, star formation, stellar mass or central densities, the correlation of sizes with environment is always inexistent.

\end{itemize}

Several recent works have studied the effect of environment on the mass-size relation of massive early-type galaxies in the local universe. \cite{2010MNRAS.402..282M} reported no size difference between ETGs living in nearby clusters ($z<0.4$) and those residing in the field  with similar stellar mass. In the same line, \cite{2013MNRAS.434..325F} studied a sample of isolated ETGs in the SDSS and found no difference in size when compared with less isolated galaxies. \cite{2012arXiv1211.1005P} also tackled this issue with an independent sample and found a trend (though not statistically significant, i.e. $1\sigma$) pointing towards smaller galaxies in clusters. Given that the difference is not statistically significant it should not be considered as a real discrepancy with the present work. We have also shown that the small difference might be a consequence of morphological selection. All these results, seem to converge towards a picture in which the effect of environment in the structure of nearby ETGs at low redshift is neglectable. 

If the recent results pointing towards larger sizes of ETGs in cluster environments at $z>1$ are confirmed (e.g. \citealp{2013arXiv1307.0003D, 2012ApJ...750...93P, 2013arXiv1307.3247L}), our results suggest that the effect has to disappear between $z\sim1$ and $z\sim0$. We will reserve a full theoretical study of current predictions from hierarchical models in a companion paper (Shankar et al., in prep.) in which we will show how environment can be used to put constraints on the physical processes behind mass assembly of early type galaxies. Overall, a variety of parameters contribute to the environmental dependence of sizes in the models, from the exact choice of dynamical friction timescale, to the level of stripping in merging satellites. Clearly, a detailed investigation of this is beyond the scope of the present work and will be presented in a dedicated work.

%A full theoretical work describing the implications of these results for the hierarchical growth of ETGs will follow this work (Shankar et al. in preparation). In these models, bulges grow mainly via two mechanisms: mergers and disk instabilities. Mergers contribute essentially to build elliptical galaxies with large bulges ($B/T>0.7$) while disk instabilities are more important in galaxies with intermediate B/T=0.4-0.7. 

FS acknowledges support from a Marie Curie grant. MB, AM and VV acknowledge support from NASA grant
ADP/NNX09AD02G and NSF/0908242. We also thank I. Trujillo, G. Mamon and B. Poggianti for interesting discussions.




%\bibliographystyle{aa}
%\bibliography{biblio_MNRAS}

\clearpage

\end{document}